\documentclass[twoside]{article}

\usepackage{PRIMEarxiv}

\usepackage[utf8]{inputenc} 
\usepackage[T1]{fontenc}    
\usepackage{hyperref}       
\usepackage{url}            
\usepackage{booktabs}       
\usepackage{amsfonts}       
\usepackage{nicefrac}       
\usepackage{microtype}      
\usepackage{subcaption}
\usepackage{lipsum}
\usepackage{svg}
\usepackage{fancyhdr}       
\usepackage{graphicx}       
\usepackage{siunitx}
\graphicspath{{media/}}     
\usepackage{cancel}
\usepackage{multicol}
\usepackage{hyperref}
\usepackage{booktabs}
\usepackage[utf8]{inputenc}
\usepackage[cmex10]{amsmath}
\usepackage{algorithm}
\usepackage{algorithmic}
\renewcommand{\figurename}{Fig.}
\renewcommand{\tablename}{Table}

\newcommand{\figref}[1]{\textcolor{blue}{\figurename~\ref{#1}}}
\newcommand{\tabref}[1]{\textcolor{blue}{\tablename~\ref{#1}}}

\hypersetup{
    colorlinks=true,
    linkcolor=blue,
    citecolor=blue,
    filecolor=blue,
    urlcolor=blue
}

\usepackage{multirow}

\title{Small-Signal Analyses Using Analytical IBR Models and Frequency-Dependent Thévenin Equivalents
\thanks{\textit{\underline{\textbf{Accepted} at 5th International Conference on Power Systems and Electrical Technology, Osaka}} \\} 
}

\author{
Nicolae Darii \\
Siemens Gamesa Renewable Energy A/S \\
Technical University of Denmark \\
Kgs. Lyngby, Denmark \\
\texttt{nidar@dtu.dk}
\And
Alberto Magagna \\
Siemens Gamesa Renewable Energy A/S \\
Kgs. Lyngby, Denmark \\
\texttt{alberto.magagna@siemens-energy.com}
\And
Oscar Sabor\'{i}o-Romano \\
Technical University of Denmark \\
Kgs. Lyngby, Denmark \\
\texttt{email@email}
\And
Nicolaos A. Cutululis \\
Technical University of Denmark \\
Kgs. Lyngby, Denmark \\
\texttt{email@email}
}

\begin{document}
\maketitle

\begin{abstract}
This paper investigates whether component-level studies can capture additional interactions through Small Signal Analysis (SSA) when the network connected to the Voltage Source Converter (VSC), typically modeled as a simple Thévenin Equivalent, is a more complex IBR-based network. The research investigates cases ranging from basic analytical to an IEEE 9-Bus EMT model, with and without Inverter-Based Resources (IBRs), synthesized as State-Space elements. The study identified that spurious poles at $50 Hz$ related to $dq$-frame conversion can hinder the accuracy of participation factor analysis. A potential approach involves a two-step process: first, applying Henkel reduction to remove most spurious poles, followed by manual elimination of any remaining ones.
\end{abstract}


\section{Introduction}
The power system is shifting toward a new paradigm thanks to the introduction of Inverter-Based Resources (IBR), namely renewables. The network structure, as well as the methods used to study and design it, require a paradigm shift \cite{Cheah-Mane2026AElectronics}. Traditionally, the electro-mechanical nature of devices such as synchronous generators permitted the use of Root Mean Square (RMS) based analysis, which is efficient for slow transients and steady-state system-level studies. However, the introduction of IBRs into the power system has shifted the phenomenon to different frequencies. Therefore, it necessitates moving towards Electromagnetic Transients (EMT) studies, based on numerical solutions of Differential Algebraic Equations (DAE), which can capture a wider range of phenomena at a faster rate, as is typical of IBRs. Moving forward, when additional information on dynamic stability or quantification of contributions is needed, the Small-Signal Analysis (SSA) can provide global information about the system under study.

To conduct SSA studies, it is necessary to model the system as a DAE. 
As for a single IBR connected to a simple Thevenin Equivalent, when its internal structure and tunings are also known, the SSA is straightforward \cite{Milano2010PowerScripting}.
When the study aims to interface the IBR to a more complex system, it would require full knowledge of the other system's components. This is where the IBR's presence, characterized by Intellectual Property (IP) protection \cite{FosteringUNECE}, hinders the ability to model the system in DAE form. For this reason, when it is required to perform stability analyses considering how a known IBR would interact with the system, Impedance-Based methods can be implemented \cite{Mugambi2025MethodologiesAnalysis}. However, this method can furnish limited information comparable to a full SSA analysis, which can be performed by considering a known IBR connected to a simplified version of the power system in Thevenin Equivalent form. This aspect raises the question of whether the Thevenin equivalent generator simplification, commonly used even in analyses that require precision, such as SSA, remains valid when studying larger systems that include IBRs. 

In the literature, attempts have been made to perform system-level SSA directly from frequency-response data extracted via a frequency scan. The frequency response can provide important information about passivity, the most undamped frequency, and the phase margin \cite{Cifuentes2022Black-BoxGrid}. It can also provide information on the eigenvalues, yielding an almost complete SSA study. For this, it requires, in addition to the frequency response extraction, also the usage of vector fitting techniques to synthesize a numerical state-space characterized by (almost) an arbitrary, unlabeled number of states \cite{Garcia-Reyes2025Data-DrivenConverters}. The Thevenin Equivalent generator is usually defined at the nominal frequency, based on the Short Circuit Ratio (SCR) and XR ratio of the grid it is trying to simplify. An attempt to build a more complex Thevenin Equivalent, that considers a wider frequency range, can lead to a Frequency Dependent Thevenin Equivalent \cite{Rogalla2020DeterminationSpectroscopy}.

For this reason, this research aims to combine different methods associated with SSA analysis to verify the limits of the Thevenin Equivalent generator model, while performing an SSA on a fully known IBR model connected to a gradually more complex network. In order to do this, the research aims to compare the SSA analysis performed earlier with an analytical Thevenin equivalent model with different Frequency Dependent Thevenin Equivalent models, synthesized in state space form, from networks (from a simple Thevenin equivalent to IBR based IEEE 9-Bus \cite{IEEEPSCAD}) built and scanned in an EMT software.

\section{Methodology}
\label{Sect: mathod}
This section provides an overview of the methods used to perform the stability studies with a hybrid analytical-numerical approach.
When SSA is performed on VSC converters connected to a Thévenin equivalent generator model, the study can be conducted fully analytically. However, the goal is to explore potential additional interactions via SSA when the system connected to the VSC in the study is more complex and also consists of IP-protected IBRs. 

The required mathematical tools for such a task will be: State-Space (SS) models, thanks to the connected modal analysis and connection method properties, Vector Fitting (VF), since it is necessary for the extraction of SS models when analytical form is not available, and Frequency Sweep (FS) to extract frequency responses (FRs). Listed here for clarity:

\begin{itemize}
    \item Analytical State-Space models (SS)
    \item Vector Fitting (VF)
    \item Frequency Sweep (FS)
\end{itemize}


\subsection{Analytical State Space models}
\begin{figure}[b!]
    \centering
    \includegraphics[width=1\linewidth]{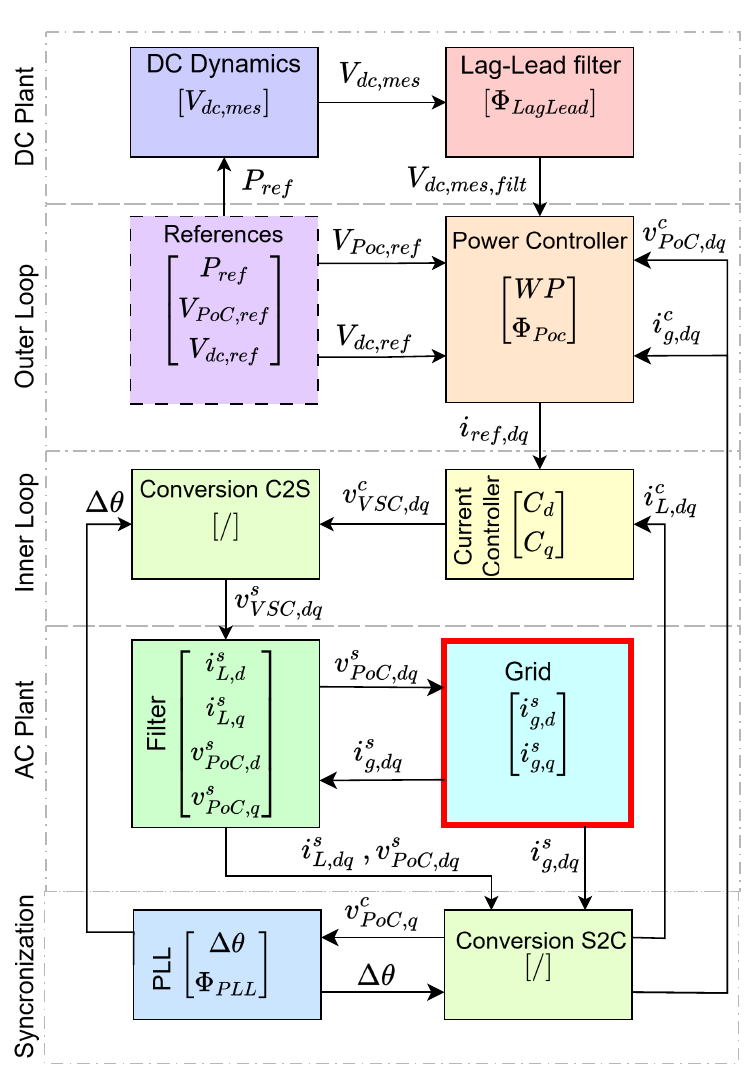}
    \caption{VSC State Space Structure}
    \label{fig:VSC_SS}
\end{figure}

The analytical modeling of the VSC converter can be defined by a complete set of linearized differential equations that describe the controls and the electrical elements of the system. Explicitly express it in state-space form, with clearly defined I/O and state labels, as shown in \figref{fig:VSC_SS}. This form can be used in a modular fashion via the Component Connection Method \cite{Hou2017HarmonicMethod}.

Generally, the SSA can be performed immediately on the analytical state-space if the network is modeled as a Thévenin equivalent, as shown in \eqref{eqn:th-grid}. 

\begin{equation}
\begin{aligned}
\label{eqn:th-grid}
\dot{i_{g,d}^s} &= \frac{\omega_b}{L_{g_{pu}}} \left( v_{PoC,d}^s - v_{g,d}^s - R_{g_{pu}} i_{g,d}^s + \omega L_{g_{pu}} i_{g,q}^s \right) \\
\dot{i_{g,q}^s} &= \frac{\omega_b}{L_{g_{pu}}} \left( v_{PoC,q}^s - v_{g,q}^s - R_{g_{pu}} i_{g,q}^s - \omega L_{g_{pu}} i_{g,d}^s \right)
\end{aligned}
\end{equation}

The next step is the replacement of the grid elements in \figref{fig:VSC_SS} (red box) with an SS element derived from a more complex IBR-based network. 

\subsection{Vector Fitting}

Alternatively, when analytical models are unavailable, it is still possible to perform SSA using numerical state-space models derived from FR datasets \cite{Garcia-Reyes2025Data-DrivenConverters, Haugaard2024Immittance-basedComponents}. It is possible to synthesize, through VF  \eqref{eqn: VF-MIMO}, where the numerical SISO state-space, ${H}_{ij}(s)$, is identified from a FR. Firstly, the number of poles ${N}_{ij}$ is chosen. Then, through an optimization process, the residues ${R}_{ij}$, that make the numerical SISO transfer function fit the FR dataset are extracted. Hence, the different SISOs extracted can be combined to form a Multi-SISO, which, in its final form, yields a MIMO numerical state-space. 

\begin{equation}
\label{eqn: VF-MIMO}
    {H}_{ij}(s) = {R}_{ij,0} + \sum_{l=1}^{N_{ij}}{\frac{{H}_{ij,l}}{s-{p}_{ij,l}}} \quad i,j=1,...,P
\end{equation}

This method is particularly effective for grid-connected VSC converters, as it is generally formulated as an MIMO state-space model in the $dq$ frame. 

When the SSA is performed on the final numerical state-space derived via VF, it is possible to perform the modal analysis; however, it is not possible to decompose the effects, as performing a Participation Factor (PF) analysis would result in a set of unknown states. In this regard, the literature distinguishes two cases:

\subsubsection{Closed-Loop frequency response}
When the FR is directly extracted from the I/O relationship of a closed-loop system, the SSA yields the eigenvalues of the grid-converter system \cite{Garcia-Reyes2025Data-DrivenConverters}, with the aforementioned peculiarity of not allowing meaningful information to be collected from PF studies.

\subsubsection{Open-Loop frequency responses}

\begin{figure}[h]
    \centering
    \includegraphics[width=0.7\linewidth]{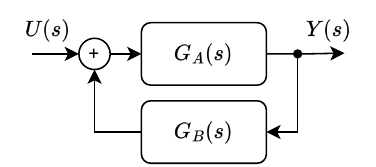}
    \caption{VF Open Loops Concatenation}
    \label{fig:OpenL}
\end{figure}

Meanwhile, when the FR is extracted through VF from two separate systems $G_A$ and $G_B$, it is possible to concatenate them, as shown in \figref{fig:OpenL}, through a feedback loop to compose the whole system. This method allows us to distinguish the states belonging to system $A$ from the $B$ even once concatenated in the final numerical state space representation \cite{Haugaard2024Immittance-basedComponents}. Although it is possible to distinguish the states' system origins, it remains unfeasible to obtain precise physical information within the $A$ and $B$ systems from the PF.

\subsubsection{Model Order reduction}
Algorithms that synthesize a system's FR into state space may lead to overfitting while trying to capture as much of the dynamics as possible. Therefore, it may use additional poles $p_{ij}$ with negligible residue terms $H_{ij}$ that are not representative of the internal dynamics but are mathematical artifacts that help achieve the fitting objective. 

Overfitting has consequences for the SSA, as in the eigenvalue analysis, it would require an additional filtering step to isolate the most significant modes. This is even more relevant for participation factor studies, which may collect contributions from the nonphysical mathematical artifacts, totally misleading the study. 

It is then possible to use a filtering technique based on Henkel singular values that identifies the minimum-order state space by evaluating the number of states that hold the minimum amount of energy \cite{Haugaard2024Immittance-basedComponents}. The procedure is:

\begin{itemize}
    \item Compute the Henkel singular values \(\sigma_i = \sqrt{\lambda_i(W_o W_c)}\) (where $W_o$ and $W_c$ are the controllability and observability gramians) and sort them in descending order: \(\sigma_1 \geq \sigma_2 \geq \cdots \geq \sigma_n\).
    \item Choose the smallest \(r\) such that \(\displaystyle\sum_{i=r+1}^{n} \sigma_i^2 \leq \epsilon\). Where $\epsilon$ is the selected tolerance for the energy.
    \item Evaluate, either graphically or quantitatively, whether \(r\) can be reduced further without exceeding a given tolerance.
\end{itemize}

\subsection{Frequency Sweep}
The FR of a component can be obtained through a Frequency Sweep, where it is excited at the input(s) and the response is observed at the output(s) at each frequency. Then, the FR $G(j\omega_k)$ is composed as shown in \eqref{eqn: scan}. 

\begin{equation}
\label{eqn: scan}
    G(j\omega_k) = \frac{y_k}{u_k} = \frac{B_ke^{j(\omega_kt+\psi_k)}}{A_ke^{j(\omega_kt+\varphi_k)}} = \frac{B_k}{A_k}e^{j(\omega_kt+\psi_k-\varphi_k)}
\end{equation}

With $k$ being the excitation frequency. $A_k$ and $B_k$ being the I/O $k$-harmonic's magnitude. $\varphi_k$ and $\psi_k$ I/O $k$-harmonic's phase. 
When systems are IBR-based, SSA is performed in the $dq$ frame; therefore, the FR in \eqref{eqn: scan} is computed across multiple I/Os forming a MIMO FR \cite{RyggASystems}.

When performing a Frequency Sweep, particular attention should be paid to the operational points. The complex network extracted in FD form and later converted to SS via VF must coincide with the VSC's I/Os for the case under study. 

\section{Results and Discussion}
\label{Sect: result}
\begin{figure}[hb!]
    \centering
    \includegraphics[width=1\linewidth]{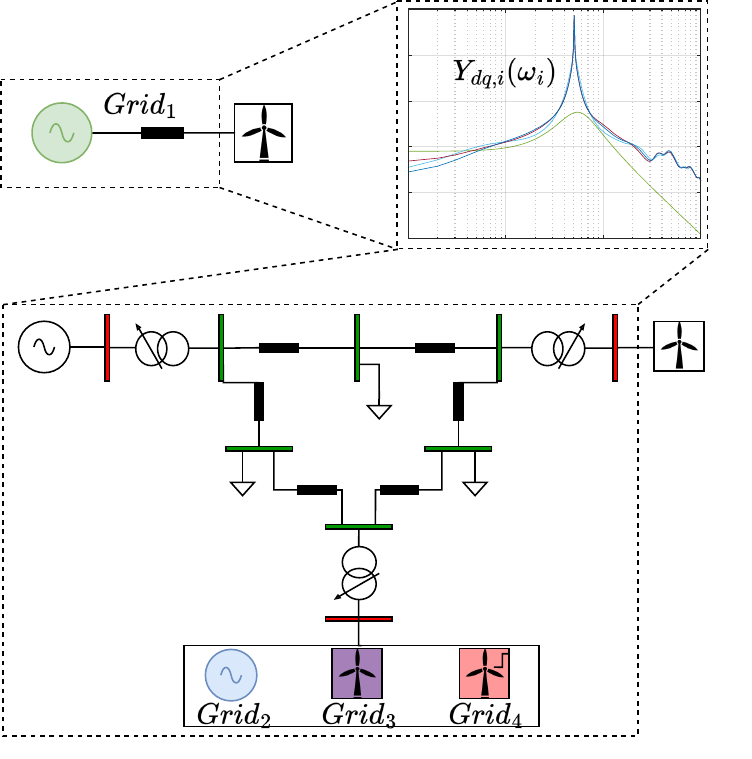}
    \caption{Frequency Scans of the different Networks}
    \label{fig:Networkscheme}
\end{figure}

    The research aims to determine whether it is possible to identify additional interactions when the network connected to the VSC is more complex than a Thevenin equivalent.
    The previous Section \ref{Sect: mathod} explained how to synthesize a complex network with IBRs into an SS version, which can later be connected to an analytical VSC SS model. 
    The current Section \ref{Sect: result} will first verify whether the synthesis method applied to an EMT-extracted version of the same analytical Thevenin equivalent in \eqref{eqn:th-grid} produces the same results. Subsequently, the application will be scaled to a more complex EMT IEEE 9-Bus network as the grid. Starting from a case without IBRs to verify whether a simple Thevenin equivalent can identify the same interactions, thus being interchangeable. Lastly, include an IP-protected IBR in the network, verify whether additional interactions are detectable, and determine the principal VSC's participation in the study. All the cases are shown in \figref{fig:Networkscheme} and listed here:

\begin{enumerate}
    \item Simple Thévenin equivalent scanned and vector fitted from an EMT model (Grid 1)
    \item IEEE 9 Bus system, without IBR, scanned and vector fitted from EMT model (Grid 2)
    \item IEEE 9 Bus system, with IBR in normal operation, scanned and vector fitted from EMT model (Grid 3)
    \item IEEE 9 Bus system, with IBR saturated, scanned, and vector fitted from EMT model (Grid 4)
\end{enumerate}

Every frequency-dependent Thévenin equivalent, synthesized in state-space form, is connected to an analytical VSC converter as shown in \figref{fig:VSC_SS}. Thus, the SSA was performed to compare the different grid complexities and assess the pros and cons.

\begin{figure}[h]
    \centering
\includegraphics[width=1\linewidth]{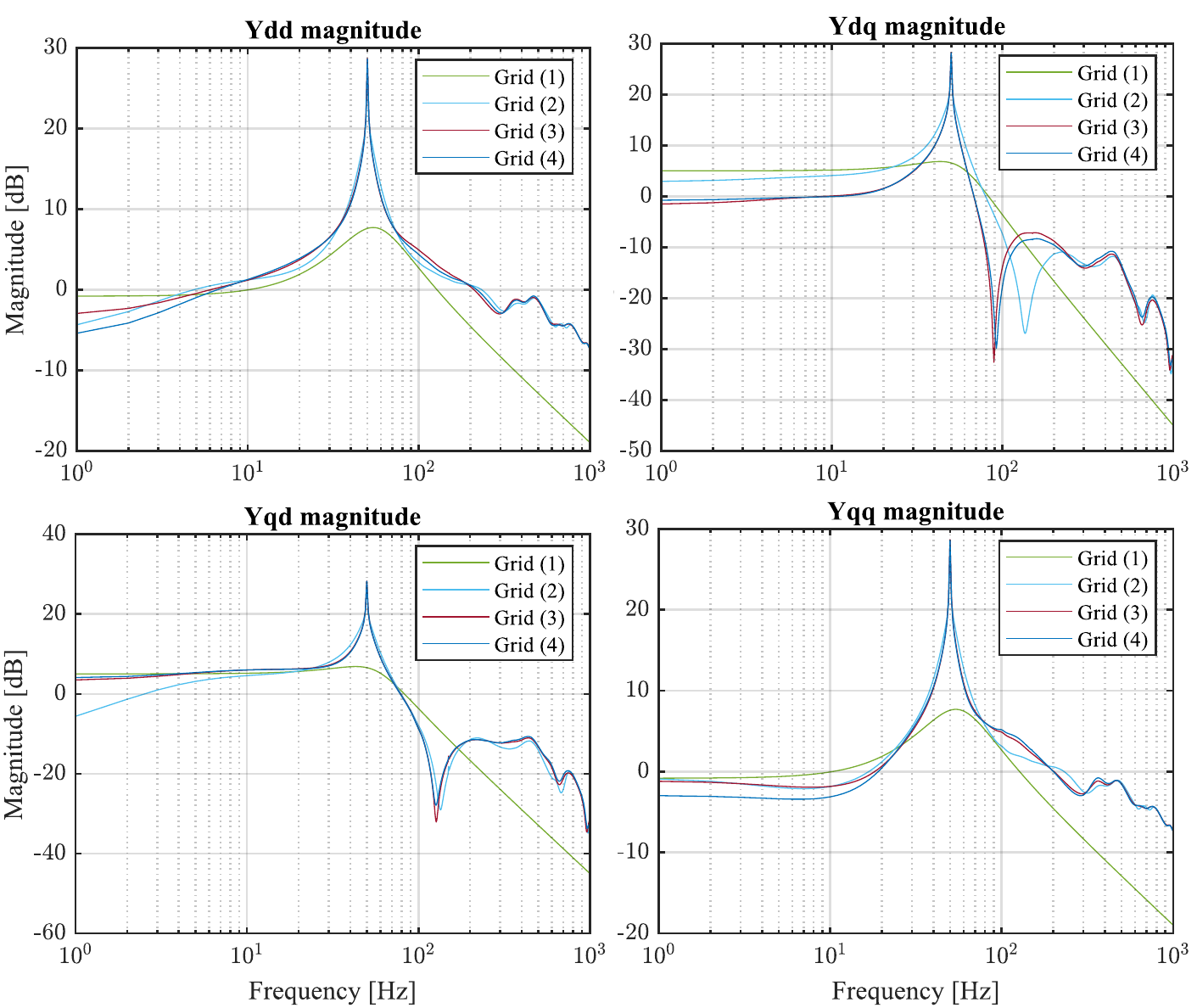}
    \caption{Networks setup in EMT}
    \label{fig:Grids}
\end{figure}

Grid 1 is designed to match the SCR and X/R of the IEEE 9Bus without IBRs.
Every scenario was scanned, as shown in \figref{fig:Grids}, in order to obtain the frequency response of each case in the form of equivalent admittance $Y_{dq,i}(\omega_i)$. The scanning range was selected as $1-1000Hz$, which results in a long enough range to take into account the worst case where the connections are modeled as simple $RL$ branches \cite{DArco2019Time-Invariant-sections}. 

\subsubsection{Validation of simple Frequency dependent Thévenin equivalent}
The simple frequency-dependent Thévenin equivalent extracted in EMT can serve as a benchmark for the method, which is later extended to more complex networks, since it can be easily reproduced analytically as in \eqref{eqn:th-grid}. When the frequency response of Grid 1, visible in \figref{fig:Grids}, is vector fitted with a minimum of $2$ poles per I/O couple, it will produce a state-space of $8^{th}$ order. Thus, by applying the Hankel reduction (poles holding $99\%$ energy), the system is reduced to a $2^{nd}$- order system, consistent with the number of states in \eqref{eqn:th-grid}. The comparison among the eigenvalues location of: the analytical Thévenin equivalent, non-reduced Simple Frequency dependent Thévenin equivalent, and its reduced form is shown in \figref{fig:Th}.

\begin{figure}[h]
    \centering
    \includegraphics[width=1\linewidth]{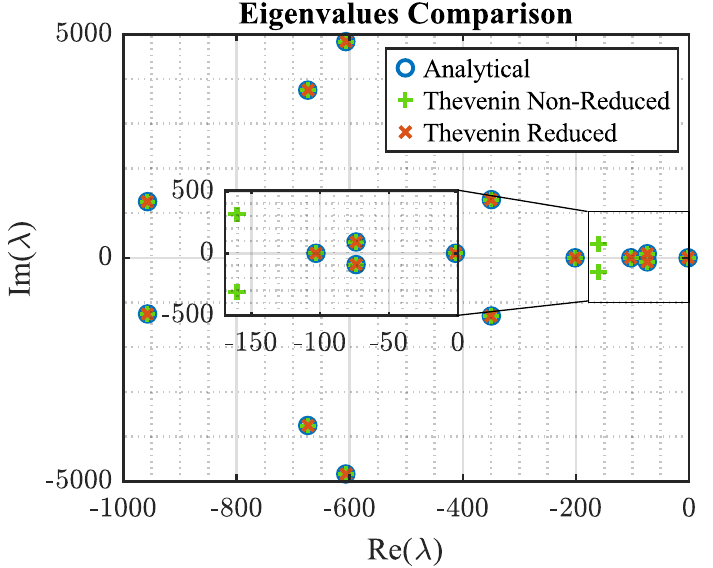}
\end{figure}

\paragraph{Spurious poles in vector fitting}
It is possible to notice additional poles in the non-reduced case, which are obviously present because of the overfitting of the VF practice. A notable aspect is the presence of multiple spurious poles around $ 50 Hz$. This is because the VF interprets this peak as a resonance point with critically low damping. While in a positive sequence, the admittance would have a simple pole as shown in \eqref{eqn: Ysimple}. 

\begin{equation}
    Y(s) = \frac{1}{R + sL}, \qquad \text{pole: } s = -\frac{R}{L}
    \label{eqn: Ysimple}
\end{equation}

The source of the resonance, visible in any single case in \figref{fig:Grids} is due to the coupling term from the $dq$ frame conversion. When inverting the $Z_{dq}$, the determinant of the matrix appears at the denominator \eqref{eqn: Ydq}. Thus, studying the resultant poles in \eqref{eqn: poles Ydq} shows a resonant pole couple with a peak, depending on the $R/L$ ratio.

\begin{equation}
\label{eqn: Ydq}
Y_{dq}(s) = \frac{1}{(R + sL)^2 + (\omega_0 L)^2}
\begin{bmatrix}
R + sL & \omega_0 L \\
-\omega_0 L & R + sL
\end{bmatrix}
\end{equation}

\begin{equation}
\label{eqn: poles Ydq}
s_{1,2} = -\frac{R}{L} \pm j\omega_0
\end{equation}

Therefore, the resonant poles that appear only in $dq$ are primarily due to Park's mathematical artifact. This influences how the VF places the poles and how they are reflected in the final closed loop. 

\paragraph{SSA with simple Frequency Dependent Thévenin equivalent}
Regarding the most undamped pole, it is consistent across both the Thévenin-reduced and non-reduced cases shown in \tabref{tab:modal_critical}.

\begin{table}[h!]
\centering
\caption{Modal analysis: Critical mode features.}
\label{tab:modal_critical}
\begin{tabular}{l c c}
\hline
Case & Frequency (Hz) & Damping (\%) \\
\hline
Base & 769.95 & 12.447 \\
Thévenin reduced & 770.19 & 12.434 \\
Thévenin non reduced & 770.2 & 12.436 \\
\hline
\end{tabular}
\end{table}

The most significant aspect of the modal analysis, once performed, is the participation factor analysis shown in \tabref{tab:participation_factors}.
The non-reduced results in a complete mismatch of the states compared to the base version, as the non-physical additional poles incorrectly take on the share of responsibility, making the whole analysis impractical.
While in the reduced case, the VSC's states match, and it is clearly deducible that $Th_1=i_{gq}^s$ and $Th_2=i_{gd}^s$.

\begin{table}[h!]
\centering
\caption{Participation factor benchmark case - 770 Hz pole}
\label{tab:participation_factors}
\begin{tabular}{l c c c}
\hline
State & Base & Thevenin reduced & Thevenin non-reduced \\
\hline
1° & $i_{Lq}^s$ (34.35\%) & $i_{Lq}^s$ (33.57\%) & $Th_1$ (33.81\%) \\
2° & $i_{Ld}^s$ (27.65\%) & $i_{Ld}^s$ (27.02\%) & $\Phi_{PoC}$ (20.43\%) \\
3° & $i_{gq}^s$ (14.56\%) & $Th_1$ (18.38\%) & $Th_4$ (20.13\%) \\
4° & $i_{gd}^s$ (14.06\%) & $Th_2$ (17.72\%) & $Th_3$ (19.55\%) \\
5° & $C_q$ (4.02\%) & $\Phi_{PoC}$ (1.85\%) & $\theta$ (1.99\%) \\
\hline
\end{tabular}
\end{table}

\subsubsection{Extension to 9 Bus IEEE Cases}

\begin{figure}[h]
    \centering    \includegraphics[width=1\linewidth]{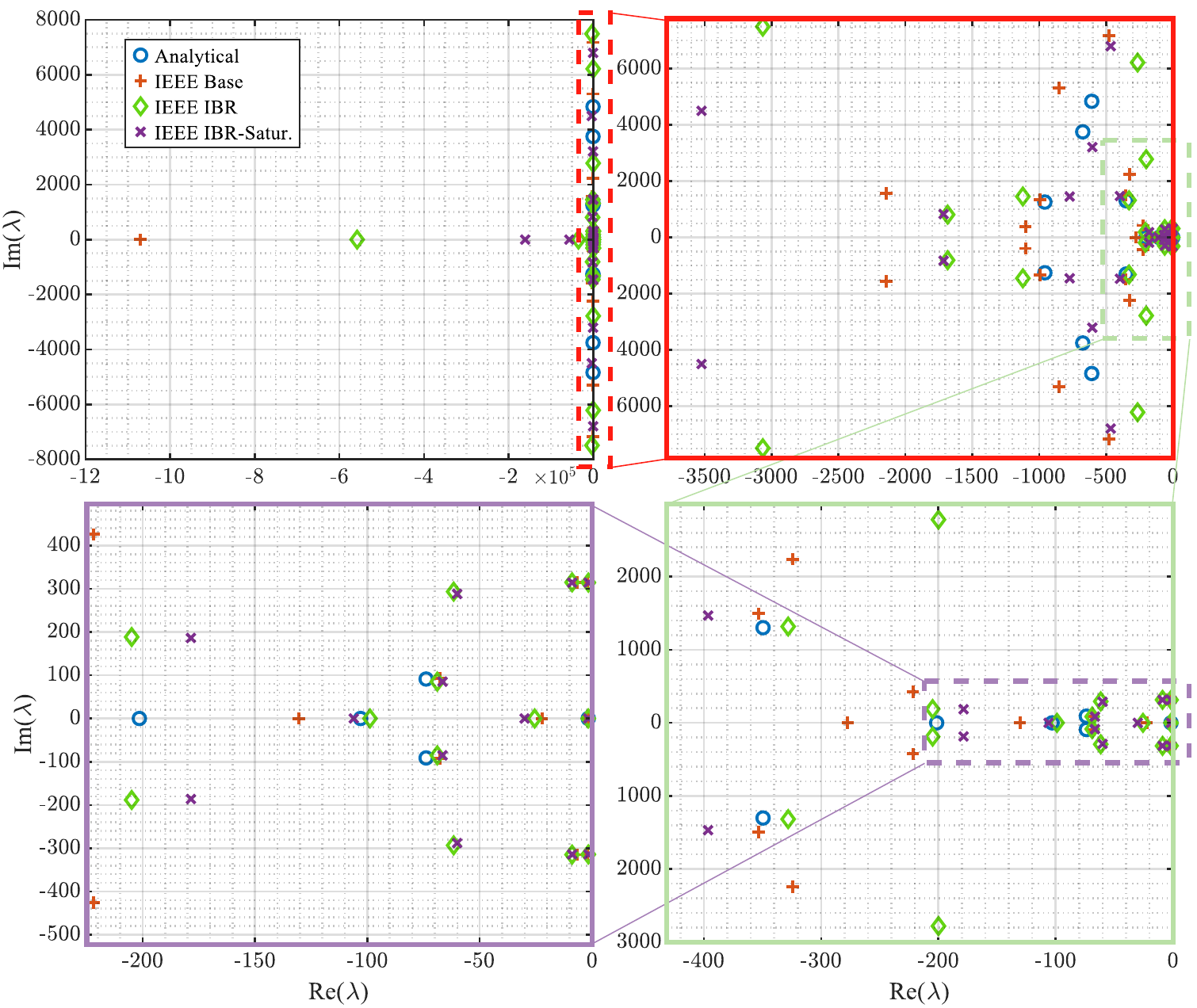}
    \caption{Thévenin and IEEE-cases Eigenvalues Comparison}
    \label{fig:Full}
\end{figure}

After verifying the combination of the analytical VSC with a synthesized (simple) network from EMT software, the research examines the scalability to more complex network structures, such as the 9 Bus IEEE case (with and without IBRs), to assess the potential and limitations. The eigenvalues of the IEEE Networks + VSC are shown in \figref{fig:Full} and natural frequencies (in $Hz$) and damping (in \%) accessible in \tabref{tab:eigenvalue_frequencies}.

\begin{table}[h!]
\centering
\footnotesize
\setlength{\tabcolsep}{3pt}
\caption{Eigenvalues frequencies and damping by mode [Hz,(\%)]}
\label{tab:eigenvalue_frequencies}
\begin{tabular}{lllllll}
\hline
Mode & Base & \textbf{$G_1$} & \textbf{$G_{1,red}$} & \textbf{$G_{2,red}$} & \textbf{$G_{3,red}$} & \textbf{$G_{4,red}$} \\
\hline
1-2 & 770 (12) & 770 (12) & 770 (12) & 1142 (7) & 1192 (38) & 1081 (7) \\
3-4 & 597 (18) & 597 (18) & 597 (18)& 844 (16) & 989 (4)& 716 (62)\\
5-6 & 207 (26) & 207 (26) & 207 (26)& 356 (14) & 442 (7)& 511 (18)\\
7-8 & 200 (61) & 200 (61) & 200 (61)& 249 (81) & 231 (61)& 234 (26)\\
9-10 & 15 (63)& 50 (45) & 15 (63)& 237 (23)& 210 (24)& 231 (47)\\
11-12 & 0 & 50 (45)& 0 & 214 (60) & 129 (90)& 131 (90)\\
13-14 & 0,/ & 50 (46)& 0,/ & 68 (46)& 50 (3)& 50 (0.5)\\
15-16 & / & 15 (63) & / & 62 (94) & 50 (0.5)& 50 (3)\\
17-18 & / & 0 & / & 50 (2)& 47 (21)& 46 (20)\\
19-20 & / & 0,/ & / & 50 (0.7)& 30 (74)& 30 (70)\\
21-22 & / & / & / & 15 (60)& 14 (63)& 14 (62)\\
23 & / & / & / & 0 & 0 & 0 \\
24 & / & / & / & 0 & 0 & 0 \\
25 & / & / & / & 0 & 0 & 0 \\
26 & / & / & / & 0 & 0 & 0 \\
27 & / & / & / & 0 & 0 & 0 \\
\hline
\end{tabular}
\end{table}

\paragraph{9Bus-IEEE without IBRs ($G_2$)}

The Grid 2 case, the benchmark case without IBRs, already shows differences in the frequency response, as shown in \figref{fig:Grids}. There is still the $50Hz$ resonance due to \eqref{eqn: Ydq}, although far more pronounced due to a low $R/L$ ratio in \eqref{eqn: poles Ydq} and additional poles $>100Hz$. The behavior is symmetrical along $Y_{dd,qq}$ and $Y_{dq,qd}$. In the eigenvalue analysis, additional modes are observed compared to the analytical and the basic reduced simple Thévenin models, as shown in \tabref{tab:eigenvalue_frequencies}. The effect of the spurious $50Hz$ poles is particularly evident in modes $\{17,18,19,20\}$. Setting aside the spurious poles, the divergence from the base case in the modes above $50Hz$ and below $1000Hz$ indicates that a complex network without IBRs cannot be replaced by a Thevenin equivalent.

\paragraph{9Bus-IEEE with IBRs ($G_3$,$G_4$)} 
In the Grid 3 and Grid 4 cases, with respective IBRs, normal and saturated, do not present significant differences compared with the Grid 2 case in the frequency response, except for an asymmetrical $Y_{dq,qd}$ response. More particularly, in the $Y_{dq}$ case, the $\sim 100Hz$ resonant peak is shifted and more pronounced. The number and magnitudes of the eigenvalue frequencies do not differ significantly from those in the Grid 2 case. In this case, the $\{13,14,15,16\}$ low-damping modes are present at $50Hz$.

\subsection{Observations Derived from Key Results}

The inclusion of a more complex grid-state-space model derived from measured systems into an analytically detailed converter model is possible. It could allow for deeper SSA analysis for component-level studies, such as tuning and precise allocation (at the variable level) for potential interactions. However, the advantage of using a complex grid-state-space model over a simple Thévenin equivalent extracted using a frequency scanner is not obvious. This is caused by the need for a spurious-modes filtering criterion to prevent biased participation-factor analysis.

Modal analysis can also be successfully performed on more complex networks. The final result would still contain some spurious modes, which can be neglected in the analysis of the most relevant modes. However, such analysis is not the key aspect for component-level studies, where the participation factor analysis also matters. 

The most significant hindrance to implementing hybrid analytical-measured elements is the reduction in the "real" minimum number of states of the unknown extracted network. Otherwise, it is not possible to perform a truthful and meaningful PF analysis, which is the main reason for this hybrid implementation in the first place.

The research also identified that the frequency-dependent admittance in $dq$ exhibits a non-physical resonant peak at $ 50 Hz$ due to the Park-transform shift.
Using the Henkel-reduction, with different levels (from least to most restrictive), the $50Hz$ redundant poles were the last to be removed. In the simple Thévenin case, the complete removal of these modes led to a successful SSA; however, in more complex cases, Henkel was unable to remove them, potentially due to differences in the resonance peaks (and thus the energy involved). A potential fix for this is to use a two-step method. First, using the Henkel reduction to filter most of the spurious poles, followed by a manual removal of the remaining ones.

\section{Conclusion}
This study presents a preliminary methodology for performing Small Signal Analysis (SSA) of accurate Voltage Source Converter (VSC) models connected to complex networks. Generally, accurate component-level SSA studies require modeling the grid as a simple Thevenin equivalent; however, this may obscure potential interactions and the contribution of internal variables. Therefore, the study investigated the possibility of including measured-based grid elements in the state-space. The results showed that the method is practical for a simple measured Thevenin equivalent in full SSA. When extended to more complex networks, potentially including Inverter-Based Resources (IBRs), eigenvalue analysis was feasible. However, grid model order reduction proved essential to complete the participation factor analysis. Consequently, future work should focus on more targeted reduction methods. The research also identified potential criteria to determine how to sufficiently minimize the network's extracted state-space model to conduct a full SSA. 

\section{Acknowledgements}
This work is supported by the European Union as part of ADOreD project funded by the Horizon Europe MSCA programme (\href{https://www.msca-adored.eu/}{HORIZON-MSCA-2021-DN, Grant agreement 101073554})

\section{Legal Disclaimer}
Figures and values presented in this paper should not be used to judge the performance of Siemens Gamesa Renewable Energy technology as they are solely presented for demonstration purpose. Any opinions or analysis contained in this paper are the opinions of the authors and not necessarily the same as those of Siemens Gamesa Renewable Energy.

\bibliographystyle{unsrt}  
\bibliography{references}

@article{RyggASystems,
    title = {{A modified sequence domain impedance definition and its equivalence to the dq-domain impedance definition for the stability analysis of AC power electronic systems}},
    author = {Rygg, Atle and Molinas, Marta and Zhang, Chen and Cai, Xu}
}

@article{Cheah-Mane2026AElectronics,
    title = {{A New Paradigm for Small-Signal Stability Analysis in Modern Power Systems: Challenges, Models, and Methods in Power Systems Rich in Power Electronics}},
    year = {2026},
    journal = {IEEE Power and Energy Magazine},
    author = {Cheah-Mane, Marc and Arevalo-Soler, Josep and Moutevelis, Dionysios and Mateu-Barriendos, Elia and Prieto-Araujo, Eduardo and Gomis-Bellmunt, Oriol and Renedo, Javier and Martin-Almenta, Macarena and Nuno-Martinez, Edgar and Martinez-Villanueva, Sergio and Jung, Joo Yong and Kim, Namkyu and Kwon, Young Jin},
    number = {1},
    pages = {67--81},
    volume = {24},
    publisher = {Institute of Electrical and Electronics Engineers Inc.},
    url = {https://ieeexplore-ieee-org.proxy.findit.cvt.dk/document/11328961},
    doi = {10.1109/MPE.2025.3588485},
    issn = {15584216}
}

@article{Cifuentes2022Black-BoxGrid,
    title = {{Black-Box Impedance-Based Stability Assessment of Dynamic Interactions between Converters and Grid}},
    year = {2022},
    journal = {IEEE Transactions on Power Systems},
    author = {Cifuentes, Nicolas and Sun, Mingyu and Gupta, Robin and Pal, Bikash C.},
    number = {4},
    month = {7},
    pages = {2976--2987},
    volume = {37},
    publisher = {Institute of Electrical and Electronics Engineers Inc.},
    doi = {10.1109/TPWRS.2021.3128812},
    issn = {15580679}
}

@article{Garcia-Reyes2025Data-DrivenConverters,
    title = {{Data-Driven State-Space Modeling for Small-Signal Stability Analysis of Black-Box Power Converters}},
    year = {2025},
    journal = {2025 IEEE PES Innovative Smart Grid Technologies Conference Europe (ISGT Europe)},
    author = {Garcia-Reyes, Luis A. and Prieto-Araujo, Eduardo and Lacerda, Vinicius A. and Ar{\'{e}}valo-Soler, Josep and Gomis-Bellmunt, Oriol and Martin-Almenta, Macarena and Nu{\~{n}}o-Mart{\'{i}}nez, Edgar and Renedo, Javier},
    month = {10},
    pages = {1--6},
    publisher = {IEEE},
    url = {https://ieeexplore.ieee.org/document/11305456},
    isbn = {979-8-3315-2503-3},
    doi = {10.1109/ISGTEUROPE64741.2025.11305456}
}

@article{Rogalla2020DeterminationSpectroscopy,
    title = {{Determination of the Frequency Dependent Th{\'{e}}venin Equivalent of Inverters Using Differential Impedance Spectroscopy}},
    year = {2020},
    journal = {2020 IEEE 11th International Symposium on Power Electronics for Distributed Generation Systems, PEDG 2020},
    author = {Rogalla, Soenke and Kaiser, Sebastian and Burger, Bruno and Engel, Bernd},
    month = {9},
    pages = {181--186},
    publisher = {Institute of Electrical and Electronics Engineers Inc.},
    url = {https://ieeexplore-ieee-org.proxy.findit.cvt.dk/document/9244380},
    isbn = {9781728169903},
    doi = {10.1109/PEDG48541.2020.9244380}
}

@misc{FosteringUNECE,
    title = {{Fostering interoperability and open-source solutions to accelerate the digital transformation of the energy system | UNECE}},
    url = {https://unece.org/sed/documents/2025/07/working-documents/fostering-interoperability-and-open-source-solutions}
}

@article{Hou2017HarmonicMethod,
    title = {{Harmonic stability analysis of offshore wind farm with component connection method}},
    year = {2017},
    journal = {Proceedings IECON 2017 - 43rd Annual Conference of the IEEE Industrial Electronics Society},
    author = {Hou, Peng and Ebrahimzadeh, Esmaeil and Wang, Xiongfei and Blaabjerg, Frede and Fang, Jiakun and Wang, Yanbo},
    month = {12},
    pages = {4926--4932},
    volume = {2017-January},
    publisher = {Institute of Electrical and Electronics Engineers Inc.},
    url = {https://ieeexplore.ieee.org/document/8216850},
    isbn = {9781538611272},
    doi = {10.1109/IECON.2017.8216850}
}

@misc{IEEEPSCAD,
    title = {{IEEE 09 Bus System | PSCAD}},
    url = {https://www.pscad.com/knowledge-base/article/25}
}

@article{Haugaard2024Immittance-basedComponents,
    title = {{Immittance-based Black-Box Model Identification via Vector Fitting Methods for Offshore Wind Power Plant Components}},
    year = {2024},
    journal = {IET Conference Proceedings},
    author = {Haugaard, Jeppe and Malmquist, Finnur and Ghimire, Sulav and Guerreiro, Gabriel M.G. and Guest, Emerson and Yang, Guangya},
    number = {16},
    pages = {1042--1051},
    volume = {2024},
    publisher = {Institution of Engineering and Technology},
    url = {https://ieeexplore-ieee-org.proxy.findit.cvt.dk/document/10916238},
    doi = {10.1049/icp.2024.3912},
    issn = {27324494}
}

@article{Mugambi2025MethodologiesAnalysis,
    title = {{Methodologies for offshore wind power plant dynamic stability analysis}},
    year = {2025},
    journal = {Renewable and Sustainable Energy Reviews},
    author = {Mugambi, Germano Rugendo and Darii, Nicolae and Khazraj, Hesam and Sabor{\'{i}}o-Romano, Oscar and Raducu, Alin George and Sharma, Ranjan and Cutululis, Nicolaos A.},
    month = {7},
    pages = {115635},
    volume = {216},
    publisher = {Pergamon},
    url = {https://www.sciencedirect.com/science/article/pii/S1364032125003089},
    doi = {10.1016/J.RSER.2025.115635},
    issn = {1364-0321}
}

@article{Milano2010PowerScripting,
    title = {{Power system modelling and scripting}},
    year = {2010},
    journal = {Power Systems},
    author = {Milano, Federico},
    pages = {1--550},
    volume = {54},
    isbn = {9783642136689},
    doi = {10.1007/978-3-642-13669-6},
    issn = {16121287}
}

@article{DArco2019Time-Invariant-sections,
    title = {{Time-Invariant State-Space model of an AC Cable by dq-representation of Frequency-Dependent {$\pi$}-sections}},
    year = {2019},
    journal = {Proceedings of 2019 IEEE PES Innovative Smart Grid Technologies Europe, ISGT-Europe 2019},
    author = {D'Arco, Salvatore and Suul, Jon Are and Beerten, Jef},
    month = {9},
    publisher = {Institute of Electrical and Electronics Engineers Inc.},
    isbn = {9781538682180},
    doi = {10.1109/ISGTEUROPE.2019.8905577}
}

\end{document}